# Non-invasive real-time imaging through scattering layers and around corners via speckle correlations

Ori Katz[1,2*], Pierre Heidmann[1], Mathias Fink[1], Sylvain Gigan[1,2]

[1] Institut Langevin, UMR7587 ESPCI ParisTech and CNRS, INSERM ERL U979, 1 Rue Jussieu, 75005 Paris, France

[2] Laboratoire Kastler Brossel, Université Pierre et Marie Curie, Ecole Normale Supérieure, CNRS, 4 Place Jussieu, 75252 Paris Cedex 05, France

*ori.katz@espci.fr

**Imaging with optical resolution through and inside complex samples is a difficult challenge with important applications in many fields. The fundamental problem is that inhomogeneous samples, such as biological tissues, randomly scatter and diffuse light, impeding conventional image formation. Despite many advancements, no current method enables to noninvasively image in real-time using diffused light. Here, we show that owing to the 'memory-effect' for speckle correlations, a single image of the scattered light, captured with a standard high-resolution camera, encodes all the information that is required to image through the medium or around a corner. We experimentally demonstrate single-shot imaging through scattering media and around corners using incoherent light and various samples, from white paint to dynamic biological samples. Our lensless technique is simple, does not require laser sources, wavefront-shaping, nor time-gated detection, and is realized here using a camera-phone. It has the potential to enable imaging in currently inaccessible scenarios.**

Diffraction-limited optical imaging is an indispensable tool in many fields of research. It forms the basis to a variety of applications ranging from astronomical observations and satellite imagery to biomedical imaging and wafer inspection. However, unfortunately, the inherent inhomogeneity of complex samples such as biological tissues induces light scattering, which diffuses any optical beam into a complex speckle pattern[1], limiting the resolution and penetration depth of optical imaging techniques[2]. Many approaches to overcome this fundamental, yet practical, problem have been put forward over the years, with pioneering experiments in holography dating back to just a few years after the invention of the laser[3, 4]. However, to-date, no approach allows to perform noninvasive imaging in real-time using diffused light. Modern techniques that are based on using only the unscattered, 'ballistic' light, such as optical coherence tomography and two-photon microscopy, have proven very useful but are inherently limited to shallow depths where a measureable amount of unscattered photons is present. Adaptive optics techniques[5] can near-perfectly correct low-order aberrations using deformable mirrors, but require the presence of a bright point-source 'guide star' or a high initial image contrast[6]. Recent exciting advancements in controlled wavefront shaping[7] have allowed focusing and imaging through highly scattering samples[8-27]. However, these techniques either require initial invasive access to both sides of the scattering medium[8-15], the presence of a guide-star or a known object[16-19], or a long acquisition sequence that involves the projection of a large number of optical patterns onto the medium[20-27]. A recent breakthrough approach reported by Bertolotti et al. has removed the requirement for a guide-star or a prior calibration sequence by exploiting the inherent correlations in scattered speckle patterns, which are scanned over the target object[28]. However, the requirement for a set of high-resolution angular scans of a coherent laser beam[28, 29] restricts this approach to objects and scattering samples that remain completely stationary over the long acquisition process.

Here, we present a technique that allows to non-invasively image hidden objects through visually opaque layers and around corners in real-time, using incoherent light and a standard digital camera (Fig.1a-b). We show that a single image of the scattered incoherent light that diffuses through a scattering medium encodes all of the information that is required to image through it, with diffraction limited resolution. Specifically, derived from concepts used in 'stellar speckle interferometry'[30-33], and the angular 'memory-effect' for speckle correlations[34-36] exploited in Bertolotti's technique[28], we show that the autocorrelation of the scattered light pattern (Fig.1c) is essentially identical to the autocorrelation of the object's image itself (Fig.1g). We then reconstruct the object's image from its autocorrelation using an iterative Fienup-type algorithm[37]. As proofs of concept, we experimentally demonstrate our single-shot noninvasive imaging technique through a variety of highly scattering samples, from optical diffusers to dynamically varying biological tissues. In addition, we demonstrate imaging 'around corners' by recording the diffuse light back-scattered off white-painted 'walls'.

Our non-invasive, single-shot technique does not require coherent laser sources, interferometric detection, raster-scanning, wavefront-shaping, nor time-gated detection, and it can be simply realized using just a camera phone, as we demonstrate experimentally.

## Results

**Principle**

A schematic of the experiment for imaging through a scattering medium, along with a numerical example is presented in Fig.1a-d. An object is hidden at a distance $u$ behind a highly scattering medium of thickness *L*. The object is illuminated by a spatially incoherent, narrowband source, and a high resolution camera that is placed at a distance *v* on the other side of the medium records the pattern of the scattered light that has diffused through the scattering medium. Although the raw recorded camera image is a low contrast, random, and seemingly information-less image (Fig.1b), its autocorrelation (Fig.1c) is essentially identical to the object's autocorrelation, as if it had been imaged by a perfect diffraction-limited optical system that had replaced the scattering medium (Fig.1e-g). The object's image is obtained from its autocorrelation by an iterative phase-retrieval algorithm[28, 31, 37] (Fig.1d).

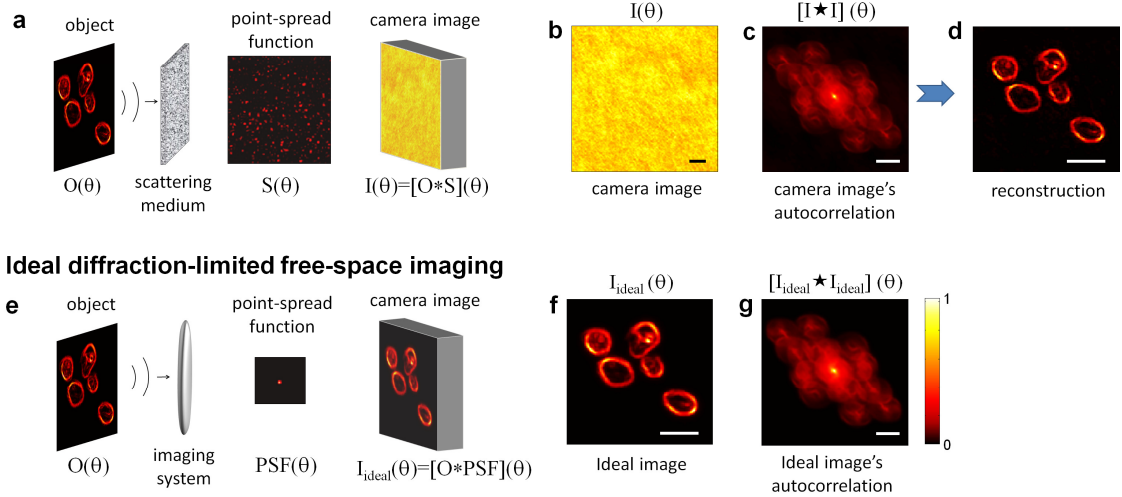

**Figure 1. Noninvasive imaging through strongly scattering layers by speckle-correlations, concept and numerical example. a,** Experimental setup: an object (a group of organelles in the numerical example) is hidden behind a visually opaque scattering medium. The object is illuminated by a spatially incoherent narrowband source, and a high resolution camera records the scattered light pattern on the other side of the scattering medium. **b,** The raw camera image, *I(θ)*. Within the memory-effect range[34], this image is given by a convolution of the object intensity pattern *O(θ)* with a random speckle pattern *S(θ)*; **c,** The autocorrelation of the seemingly information-less low-contrast camera image, *[I★I](θ)*, is essentially identical to the object's autocorrelation (g) as if it had been imaged by an aberration-free optical system replacing the scattering medium (**e-g**), as result of the sharply-peaked autocorrelation of the speckle pattern[28, 30, 31]. **d,** The object's image is obtained from the autocorrelation of (b) by an iterative phase-retrieval algorithm[28, 38]. Scale-bars are 30 camera pixels.

The reason that the autocorrelation of the diffused light is essentially identical to the object's autocorrelation is the intrinsic isoplanatism that arises from the angular memory effect for speckle correlation[34-36]. Simply put, the memory effect states that light from nearby points on the object is scattered by the diffusive medium to produce highly correlated, but shifted, random speckle patterns on the camera (Fig.1a). Points at the object plane that lie at a relative distance *Δx* that is within the memory-effect range: $Δx \ll u·λ/πL$, generate nearly identical speckle patterns on the camera. These patterns are shifted in respect to each other on the camera plane by a distance $Δy=Δx·v/u$ [18, 36]. For spatially incoherent illumination (such as fluorescence emission) no interference takes place between the different patterns, and the camera image is simply a superposition of these identical shifted speckle intensity patterns, *S(θ)* and *S(θ+Δθ)*, where *Δθ=Δx/u* is the viewing angle. This thus allows to view the system of Fig.1a as an incoherent imaging system having a shift-invariant point spread function (PSF) that is equal to this random speckle pattern, $S(\theta)$, and with a magnification of $M=v/u$. The camera image, $I(y)$, is then given by a convolution of the object intensity pattern $O(x)$ with this PSF: $I(v\theta)=O(u\theta)*S(\theta)$, where the symbol $*$ denotes a convolution operation. Taking the autocorrelation of the camera image, and using the convolution theorem yields[28, 30, 31]:

$$[I\star I](\theta) = [(O*S)\star(O*S)](\theta) = [(O\star O)*(S\star S)](\theta) \quad (1)$$

where the symbol $\star$ denotes the autocorrelation operation. As the autocorrelation of the speckle pattern, $(S\star S)$, is a sharply-peaked function[1, 28, 30] (essentially the autocorrelation of broadband noise), the right-hand-side of Equation (1) is effectively equal to the autocorrelation of the object's image $[O\star O](\theta)$. More precisely, the autocorrelation of the camera image is equal to the autocorrelation of the object as it would have been imaged by an aberration-free diffraction-limited imaging system replacing the scattering medium (Fig.1e-g), with an additional constant background term, $C$, which originates from the 2:1 peak to background ratio of the speckle autocorrelation.

$$[I\star I](\theta) = [I_{ideal}\star I_{ideal}](\theta) + C \quad (2)$$

The object's diffraction-limited image $I_{ideal}(\theta)$ can then be obtained from its autocorrelation using a Fienup-type iterative phase-retrieval algorithm[28, 31, 37] (Fig.1d, see *Methods*). The result of this procedure is that the opaque layer effectively serves as a perfect thin lens, reminding the results obtained with wavefront-controlled 'scattering lenses'[15, 18], and the cross-correlation approach of Freund[36]. However, in contrast to these approaches our technique is completely noninvasive, and it does not rely on the a-priori recording of the exact scattering properties of the sample.

Here, similar to stellar speckle interferometry[30-33] and the speckle scanning technique of Bertolotti et al.[28], the object image is recovered from its autocorrelation. However, in our technique the autocorrelation of the object is retrieved from a single image, captured in a fraction of a second, eliminating the requirement for temporal averaging over multiple images of stellar speckle interferometry, and the lengthy high resolution angular scanning of Bertolotti's approach. This breakthrough is made possible by exploiting the ergodic-like property of speckle in the diffusive regime[36], and the very high pixel-count of modern digital cameras. These allow us to recover the object's autocorrelation with high fidelity by spatially averaging the correlation function over the millions of speckle spots that are captured in a single high-resolution image, rather than performing temporal ensemble averaging over multiple exposures[30]. In fact, every camera pixel effectively plays the role of a single illumination angle in the speckle-scanning technique of Bertolotti et al.[28], parallelizing the acquisition process a million-fold.

The limit on the angular field of view (FOV) of our technique is given by the memory effect range[34], $\Delta\theta_{FOV}\approx\lambda/\pi L$. It is inversely proportional to the medium's thickness, $L$, and is, in theory, unlimited for a scattering surface of 'zero' thickness. An additional limiting factor is the finite aperture on the scattering medium that the light is collected from, which will cause vignetting and decorrelation between points that illuminate different portion of it. The effective 'entrance pupil' diameter, $D$, is dictated by the illumination solid angle and object's distance from the medium, as well as the effective diameter of any aperture stops placed between the scattering medium and the camera. This finite aperture also sets the angular imaging resolution, which is given by the diffraction limit[16, 36]: $\delta\theta\approx\lambda/nD$, where $n$ is the refractive index behind the scattering layer.

In the next sections we demonstrate the implementation of the proposed technique in a set of experiments in transmission and reflection using a variety of objects and scattering samples, and utilizing either a scientific grade sCMOS camera or a consumer-grade smartphone camera.

**Experimental imaging through visually opaque samples**

As a first experimental demonstration we used the proposed technique to image various objects through a scattering, visually opaque optical diffuser (Fig.2), and through two turbid biological samples (Fig.3). In these experiments the objects to be imaged were placed at distances of 10 to 60cm behind the scattering sample and were illuminated by a narrowband spatially-incoherent pseudothermal source (see Methods and Supplementary Figure 1). The image of the scattered light was recorded by a 5 Megapixels sCMOS camera (see Methods). The results of these experiments are summarized in Figures 2-3. The leftmost column of each figure displays the central part of the raw camera images. As can be observed, the raw camera images are low-contrast and seemingly random patterns with no visible relation to the true shape of the imaged objects. However, the autocorrelations of these seemingly-random images reveal clear distinctive patterns (the second column from the left in Figs.2-3). The hidden objects images are recovered from these autocorrelations with high fidelity by phase-retrieval[37] (see Methods). We demonstrate this reconstruction using several objects of various shapes and complexities, with different contrast of the raw camera images.

As the image of the scattered light is acquired in a single camera shot, the technique can be used to image through dynamically varying samples, such as the freshly cut shallot sample of Fig.3e-h. Interestingly, the dynamics of the sample can be exploited to yield better estimate of the autocorrelation by acquiring a set of images of the scattered light at different times (Fig.3e and Supplementary Video 1). Although the hidden object could be reconstructed from any single image of this set, averaging the autocorrelation over all 10 acquired images yields a better estimate for the autocorrelation, as result of ensemble averaging over different speckle realizations. This averaging procedure is borrowed from stellar speckle interferometry through the dynamic atmosphere[30, 31]. The natural dynamics of biological scattering samples can thus be exploited rather than fought against, to produce high quality images of hidden objects.

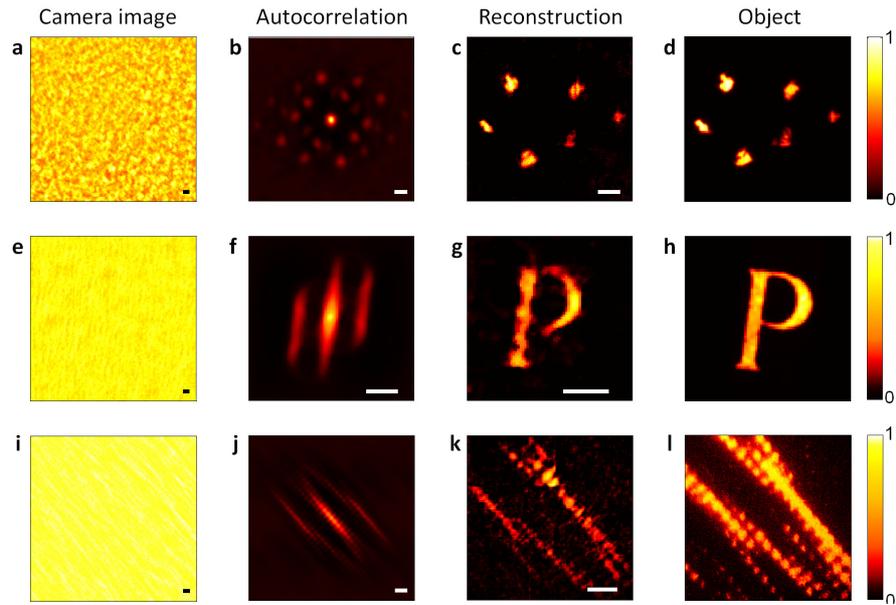

**Figure 2. Experimental imaging of various objects through a visually-opaque optical diffuser. a,** the raw camera image of the scattered light is a low contrast and seemingly random image, with no visible relation to the hidden object shape (only central part of camera image is shown); **b,** the corresponding autocorrelation of the camera image reveals clear distinctive patterns; **c,** reconstructed object's image from the autocorrelation of (b) using a phase-retrieval algorithm; **d,** The real object as imaged directly without the scattering medium. **e-f** and **i-l,** same as (a-d) for different imaged objects. Scale-bars are 20 camera pixels, corresponding to 350μm at the object plane of (a-c), 1.1mm in (e-g), and 200um in (i-k).

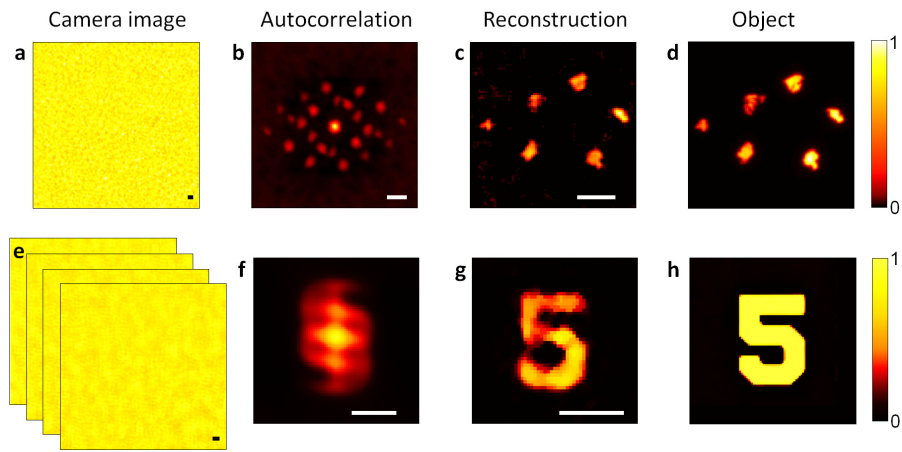

**Figure 3. Experimental imaging through biological samples: a-d,** Imaging through a ~300μm-thick chicken breast tissue; **e-h,** Imaging through fresh slice of 80μm-thick shallot skin. In e-h the relatively short decorrelation time of the dynamic sample (Supplementary Figure 2) was exploited to ensemble-average the autocorrelation over a set of 10 scattered light images taken at different times (four images are shown at the leftmost column. see supplementary Movie 1). Scale-bars are 20 camera pixels, corresponding to 700μm at the object plane of (a-d), and 450um in (e-h).

An interesting feature of the presented technique is its infinite depth of field. The target object will always be imaged with a diffraction-limited resolution regardless of its distance from the scattering layer. The reason for this is that the PSF of the diffusive light is always a speckle pattern with a sharply peaked autocorrelation function having a diffraction limited width. In contrast, a conventional lens-based system's diffraction-limited PSF has a diffraction limited autocorrelation peak only within a small axial range.

An experimental demonstration of this feature is presented in Fig.4. In this experiment we imaged a simple object composed of two bright spots at various distances from the scattering diffuser. We compare the results to those obtained when the diffuser is replaced by a conventional single-lens imaging system having an entrance pupil of the same size. As expected, the images of the object obtained through the scattering medium remain sharp whereas those taken with the conventional imaging system are sharp only when the objects are near the focus of the system. The unlimited depth of field is obtained for the same reason that extended depth of field is attained using wavefront coding techniques[39, 40]. In fact, the scattering medium is, in essence, a random wavefront coder, though unlike conventional wavefront-coding, it is an unknown and uncontrolled coder. Similar to wavefront-coding, the diffraction limited resolution and infinite depth of field are obtained at the price of a reduced image contrast, or equivalently at an expense of an increased dynamic range requirement from the imager[42, 43] (see Discussion). The downside of the extended depth of field is that in the absence of any further processing, the technique has no axial sectioning capability.

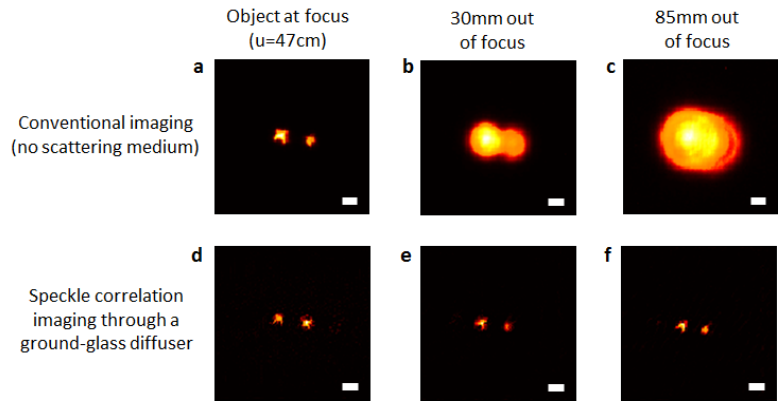

**Figure 4. Extended depth of field demonstration.** Comparison between the images of a two-point object placed at various distances, $u$, from the imaging setup, obtained with: **a-c,** a conventional single-lens imaging system, and **d-f,** a visually-opaque ground glass diffuser placed instead of the lens. Both systems have similar numerical apertures. Scale-bars are 10 camera pixels, corresponding to 300μm at the object plane of (a,d).

## Imaging 'around corners' with diffused back-scattered light

The presented technique can also work in reflection, i.e. be applied to image occluded objects 'around the corner' using the light back-scattered from a diffusive wall. An experimental demonstration of imaging using diffuse back-reflected light is presented in Fig.5. The scattering sample in this experiment is a thick layer of ZnO powder, which is essentially white paint. The imaging resolution and FOV in reflection are given by the same expressions as for transmission geometry but by replacing the medium thickness, $L$, with the transport mean-free-path $l*$[36].

Remarkably, the stronger is the scattering (shorter mean-free-path) the larger is the FOV. For common highly scattering white paint pigments such as $TiO_2$ and $ZnO$, $l^*$ of the order of one micron[8], corresponding to an angular FOV exceeding several degrees.

When imaging in reflection (Fig.5a) the scattering-medium effectively serves as a curved mirror, as envisioned by Freund[36] and demonstrated recently using wavefront shaping[18]. However, here, again, the imaging is performed completely noninvasively. The imaging is performed in real-time using only a standard camera, a significant advantage compared to other 'around-the-corner' imaging techniques[18, 41].

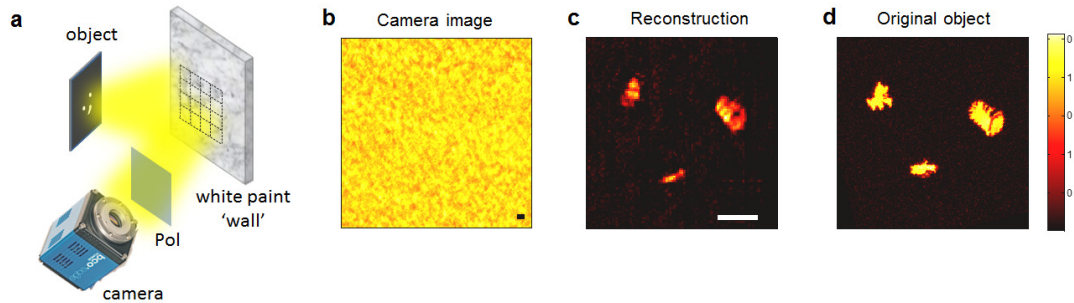

**Figure 5. Noninvasive imaging 'around corners' using back-scattered incoherent light. a,** The light from an incoherently illuminated object hits a highly scattering layer of ZnO 'white paint'. An image of the backscattered diffused light pattern at a given polarization is taken by a high resolution camera; **b,** the central part of the raw camera image; **c,** reconstructed object's image from the autocorrelation of the camera image; **d,** The hidden object's real image. Scale-bars are 20 camera pixels, corresponding to 400μm at the object plane.

**Seeing through opaque layers with a camera phone**

As a final demonstration of the simplicity and applicability of the presented technique we show that the requirements for implementing it (see Discussion) can be satisfied using a consumer-grade, high pixel-count cameras, available to millions of smartphone users today. To this end we have repeated the experiments of imaging through the visually-opaque diffuser, using a Nokia Lumia 1020 smartphone, which features a 41 Megapixel sensor. In these experiments we placed various objects 20cm from the diffuser and simply took an image of the scattered light with the camera-phone placed on the other side of the diffuser (Fig.6a-b). We retrieve the hidden objects from a single scattered light image, taken at 1/100s - 1/20s exposure times (Fig.6c-g).

The major hurdles posed by using such low-cost imaging devices are their fixed miniature optics that dictate the viewing angle and magnification, and possibly introduce aberrations that limit the FOV. In addition, the Bayer color filter sensor array introduces pixel-scale artifacts when interpolated to the full resolution, impairing the reconstruction quality. Finally, when processing images saved with the common JPEG compression and nonlinear-intensity gamma correction, additional artifacts and dynamic range limitations interfere with the reconstruction. However, these still allow the reconstruction of simple objects, as we demonstrate in Fig.6g (in Fig.6c-f we processed images captured in a linear digital RAW format).

The straightforward experiments of Fig.6 required no alignments nor any optical components, and they could be reproduced outside the lab by a non-specialist. The only necessary modification to the smartphone camera was reducing its entrance aperture, so that the individual speckle grains are resolved by more than 2 camera pixels (see Discussion). This was done by simply placing a black screen with a ~0.8mm diameter pinhole on top of the camera lens. The experimenter is then only required to "point-and-shoot" through the opaque window using his modified handheld camera.

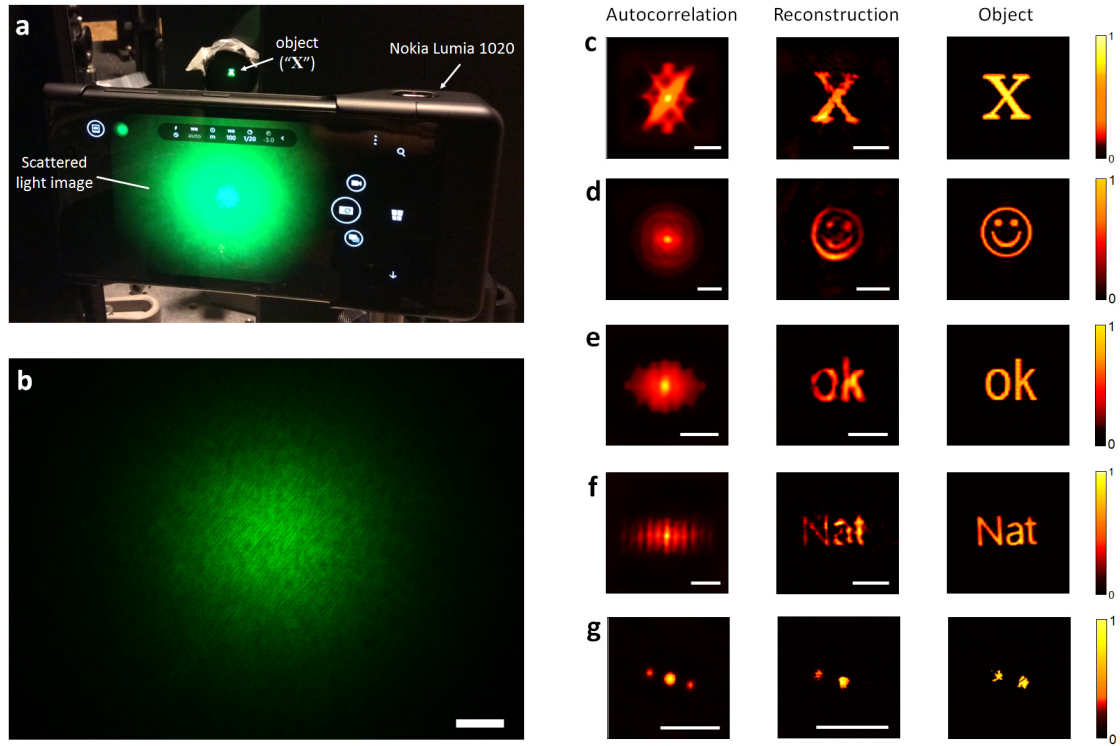

**Figure 6. Seeing through an opaque layer with a camera phone. a,** A photograph of the experiment: a camera phone (Nokia Lumia 1020) is taking a picture of the object (The letter 'X', in this image) through a highly scattering ground-glass diffuser; **b,** the raw camera image in linear intensity scale (scale-bar, 500 pixels); **c,** left column: calculated autocorrelation of the image in (b); central column: the reconstructed object from the image autocorrelation; right column: an image of the real hidden object. **(d-g)** same as (c) but for different objects. In (c-f) linear RAW format camera images were analyzed, and in (g) a non gamma-corrected JPEG image was used. Scale-bars are 50 camera pixels, corresponding to 1.7mm at the object plane.

## Discussion

Several conditions should be met in order to allow high-fidelity imaging with the presented technique: (1) The object's angular dimensions, as seen through the random medium, must be smaller than the memory-effect range (the FOV). Parts of the object that lie beyond this range would produce uncorrelated speckle patterns, and would not contribute properly to the calculated autocorrelation[29]; (2) A single speckle grain should be well resolved by the camera (adequate Shannon-Nyquist sampling), i.e. the speckle grain diameter should exceed two camera pixels. This condition is easily met by adjusting the camera distance from the scattering medium and/or by an appropriate aperture stop placed behind the scattering medium (see Supplementary Figure 1); (3) To obtain a large number of resolved resolution cells at the object plane: $B=(\Delta\theta_{FOV}/\delta\theta)^2 \propto (D/L)^2$, the available aperture diameter on the scattering medium, $D$, should be considerably larger than the medium thickness: $D>>L$. A practical limit on increasing the aperture size is the resulting reduced image contrast, which is inversely proportional to the square-root of the number of bright resolution cells (summed-up shifted speckle patterns)[29, 31]. As a consequence, the technique is more suitable to dark-field and sparsely-tagged fluorescent objects than to bright-field imaging scenarios; (4) The spectral bandwidth used to form the image must be narrower than the speckle spectral correlation width[18, 36, 42], so that the PSF, $S(\theta)$ is indeed a high-contrast speckle pattern. In other words, the detected light longitudinal coherence length should exceed the spread of optical paths traversed by the light as it diffuses through the medium. This time is given both by the Thouless time of the medium ($\Delta t \approx L^2/cl^*$ in transmission and $\Delta t \approx l^*/c$ in reflection[36]), in addition to the geometrical path lengths before and after the medium[31, 43]. This requirement for limited spectral width can be easily met by either controlling the illumination bandwidth as was done here, or by simply placing a narrowband filter in front of the camera, as was recently demonstrated with white-light illumination[18], and with fluorescent emission from an object buried in a scattering biological tissue[44]. Alternatively, several spectral bands could be processed simultaneously by separating them to several cameras, rather than filtering them out, allowing additional ensemble averaging or multispectral imaging.

The final condition is that the number of sampled and processed speckle grains should be maximized. As in other digital speckle correlation techniques, an inherent source of noise in estimating the autocorrelation function originates from replacing the ensemble averaging by averaging over a finite number of speckles (limited by the number of camera pixels)[45]. This is important for both accurately estimating the autocorrelation function as well as for correctly subtracting its constant background. In our experiments, by using modern sensors having several megapixels, we were able to obtain satisfactory SNR by processing $\sim 10^6$ speckle grains in a single camera shot. For a thorough analysis of this and other sources of noise in speckle interferometry, we refer the reader to the works of Dainty[31, 33].

In this proof of concept we have used Fienup's basic phase-retrieval algorithms[28, 37], and made no special effort to overcome the common stagnation and artifacts issues of these simple algorithms[38]. Higher fidelity reconstruction is expected by employing one of the various methods that were developed to overcome these limitations, such as the majority voting and reduced support constraint[38]. Furthermore, by employing more advanced speckle interferometry analysis techniques[31, 38], additional phase information and improved reconstruction may be obtained from the available data in some cases. These methods include triple-correlation analysis[33], Knox-Thompson[32] and exponential filtering[31, 38]. An exciting potential improvement may lie in applying the recently-developed compressive phase-retrieval algorithms[46, 47] to increase the reconstruction fidelity and reduce the acquisition time. Interestingly, if one is only interested in tracking unknown objects without reconstructing their shape, a simple cross-correlation between images taken over time can be performed[36], without requiring any phase-retrieval algorithm.

We have shown that an *unknown* random medium can serve as an imaging device, and that multiply-scattered light can be used rather than filtered out to form an image. This is closely related to the well-studied technique of time-reversal[7, 48], which has found applications in many fields[48]. A related interesting link is to passive correlation-based imaging techniques[49], where temporal correlations of broadband signals are analyzed. Similar to these techniques, our technique is lensless (the scattering medium effectively serves as the lens) and as such, it may found applications in other spectral or wave-propagation domains where fabrication of lenses is a challenge.

# Methods

## Experimental setup

The complete experimental setups for imaging in transmission and reflection are presented in Supplementary Figure 1. The imaged objects were printed transparencies, a USAF resolution target (Thorlabs R3L3S1N), and groups of holes of various shapes made on a black screen. The objects were illuminated by a narrow bandwidth spatially incoherent pseudothermal source at a wavelength of 532nm, based on a Coherent Compass 215M-50 CW laser. The light collection aperture diameter on the scattering media was limited to 0.5cm-1cm by a variable iris. The camera used in the experiments of Figs.2-5 is a PCO edge 5.5 (2560 by 2160 pixels). The camera integration time in the different experiments was 10ms to 2s (typically a few hundred milliseconds). The objects were placed at distances of 20cm to 60cm from the scattering sample and the camera was placed at distances of 9cm to 15cm from the medium. In the experiments of Fig.6 the light collection aperture was limited to ~0.9mm, and the camera-phone was placed as close as possible to the diffuser (a few millimeters distance). The optical diffuser used was a 220 grit ground-glass diffuser (DG10-220-MD Thorlabs). The ZnO sample was made by sedimentation through vertical deposition and subsequent drying of a water suspension of ZnO powder (Zinc-Oxide 205532, Sigma-Aldrich). The chicken-breast and shallot samples were placed between two microscope glass slides, which were held vertically.

## Image processing

The raw camera image intensity was spatially normalized for its varying envelope by dividing the raw camera image by a low-pass filtered version of it. The normalized image was Gaussian smoothened with a standard-deviation kernel of 0.5 to 2 pixels. The autocorrelation of the processed image was calculated by inverse Fourier transform of its energy-spectrum (effective periodic boundary conditions).

The resulting autocorrelation was cropped to a rectangular size between 100×100 pixels to 320×320 pixels (depending on the imaged object dimensions), and the minimum pixel brightness in this window was background-subtracted from the entire autocorrelation trace. A 2D Tukey window was applied on the autocorrelation before applying the phase-retrieval reconstruction algorithm in the experiments of Fig.6. The reconstructed images of Figs.6, 2a and 3c were median-filtered with a rectangular 3×3 pixels kernel.

In the camera-phone experiments of Fig.6a-f, the full 38 Megapixel resolution raw images (5360×7152 pixels) were saved in a digital negative (DNG) RAW format, and the green channel was converted to a bitmap format using DCRaw plug-in for ImageJ. The conversion was done in 16bit, linear scale, with a Variable Number of Gradient interpolation method. The converted monochrome image was used to calculate the autocorrelation as detailed above. In the experiment of Fig.6g the green channel from a full resolution JPEG format was used (without correcting the nonlinear gamma scaled intensities).

## Phase-retrieval algorithm

The phase-retrieval algorithm was implemented according to the recipe given by Bertolotti et al.[28]. Specifically, an hybrid input-output (HIO) algorithm[37] was ran with a decreasing beta factor from $\beta=2$ to $\beta=0$, in steps of 0.04. For each $\beta$ value, 40 iterations of the algorithm were performed (i.e., a total of 2000 iterations). The result of the HIO algorithm was fed as an input to additional 40 iterations of the 'error reduction' algorithm to obtain the final result. The object constraints used were real and non-negativity. The algorithms were implemented in *Matlab* and ran on an NVIDIA GeForce GTX 670 graphics processing unit (GPU) having 1344 cores, using the *gpuarray* function. A single run of all of the above iterations on this GPU took between one to two seconds. To assure faithful reconstruction of each image, several runs of the algorithm (from 20 up to 400, typically 50) were performed with different random initial conditions, and the reconstruction having the closest Fourier spectrum to the measured autocorrelation Fourier transform (lowest mean-square-error) was chosen as the final reconstructed result.

**Acknowledgements**

We thank D. Martina and A. Liutkus for help with the GPU implementation of the algorithm, P. Ducellier for the Nokia Lumia 1020 camera phone, and J. Bertolotti for helpful discussions. This work was funded by the European Research Council (grant no. 278025). O.K. is supported by the Marie Curie Intra-European fellowship for career development (IEF) and a Rothschild fellowship.

**Author contributions**

O.K. conceived the idea, performed the numerical simulations, wrote the reconstruction algorithm, and designed the initial experiments. O.K., P.H., and S.G. discussed the experimental implementation. O.K. and P.H. performed the experiments and analyzed the results. M.F, S.G., O.K discussed the results, O.K. wrote the manuscript with contributions from all authors

**Competing financial interests**

The authors declare no competing financial interests.